\documentclass[
    aps,
    prd,
    reprint,
    amsmath,amssymb,
    groupedaddress,
    twocolumn,showpacs,
    ]{revtex4-2}
\usepackage{graphicx}
\usepackage{dcolumn}
\usepackage{bm}
\usepackage[mathlines]{lineno}
\usepackage{multirow}
\usepackage{comment}
\usepackage{enumerate}
\usepackage{color}
\usepackage{amsmath}

\begin{document}


\title{Challenges in Cosmic Magnification Reconstruction by Magnification Response}


\author{Shuren Zhou $^{1,2}$}
 \email{zhoushuren@sjtu.edu.cn}
\author{Pengjie Zhang $^{1,3,2}$}
 \email{zhangpj@sjtu.edu.cn}
\affiliation{
$^{1}$ Department of Astronomy, School of Physics and Astronomy, Shanghai Jiao Tong University, Shanghai, China \\
$^{2}$ Key Laboratory for Particle Astrophysics and Cosmology (MOE)/Shanghai Key Laboratory for Particle Physics and Cosmology, China\\
$^{3}$ Division of Astronomy and Astrophysics, Tsung-Dao Lee Institute, Shanghai Jiao Tong University, Shanghai, China
}


\date{\today}


\begin{abstract}

Cosmic magnification on the observed galaxy overdensity is a promising weak gravitational lensing tracer. Current cosmic magnification reconstruction algorithms, ABS (Analytical method of Blind Separation) and cILC (constrained Internal Linear Combination), intend to disentangle the weak lensing signal using the magnification response in various flux bins. 
In this work, we reveal an unrecognized systematic bias arising from the difference between galaxy bias and the galaxy-lensing cross-correlation bias, due to the mismatch between the weak lensing kernel and the redshift distribution of photometric objects. It results into a galaxy-lensing degeneracy, which invalidates ABS as an exact solution.
Based on the simulated cosmoDC2 galaxies, we verify that the recovered weak lensing amplitude by ABS is biased low by $\sim10\%$. cILC, including a modified version proposed here, also suffers from systematic bias of comparable amplitude. Combining flux and color information  leads to significant reduction in statistical errors, but fails to eliminate the aforementioned bias. With the presence of this newly found systematic, it remains a severe challenge in blindly and robustly separating the cosmic magnification  from the galaxy intrinsic clustering.

\end{abstract}


\maketitle


\section{Introduction}
\label{sec:intro}

Over the past two decades, weak lensing has emerged as a powerful cosmological probe, and achieved great success in constraining the structure growth of the universe and nature of dark energy, neutrino mass and gravity. The CMB lensing reconstructed from the distortion in cosmic microwave background maps is the most precision weak lensing measurement to this day, with $\gtrsim 40\sigma$ detection significance \cite{aghanim2020planck, qu2024atacama}. Simultaneously in the low redshift, the galaxy lensing from measuring galaxy shapes is a powerful way to understand the late universe \cite{Hoekstra_2008, abbott2022dark, dalal2023hyper, li2023hyper, euclidcollaboration2023euclid_weaklensing, yao2023desi, xu2023desi}, and numerous works have developed effective methods to measure the cosmic shear thereby extract the weak lensing signal \cite{zhang2011measuring, zuntz2013im3shape, giblin2021kids, congedo2024euclid}. Apart from the shear component, the magnification of the galaxy lensing superimposes the foreground weak lensing signal on the observed galaxy clustering, and is a potentially powerful tracer for weak lensing reconstruction \cite{BARTELMANN2001291, Hoekstra_2008, Scranton_2005, xu2024accurate, menard2003improving, Casaponsa_2013, bauer2014magnification, bauer2012mass, Chiu_2020, Morrison_2012, Tudorica_2017, Bonavera_2021, Crespo_2022, M_nard_2010, Schmidt_2011, Jain_2011, Duncan_2016}. 

With the ongoing and upcoming surveys, such as the Euclid \cite{scaramella2022euclid}, CSST \cite{cao2018testing, gong2019cosmology, lin2022forecast} and LSST \cite{ivezic2019lsst}, the large amount of photometric samples would enable us to conduct the precision lensing measurement from galaxies. If we can develop precision methodology for cosmic magnification reconstruction from galaxy overdensity, we are expected to make full use of the abundant information from both their shapes and counts in galaxy lensing observations.
However, because the intrinsic galaxy clustering amplitude overwhelms the cosmic magnification by several orders of magnitude, disentangling cosmic magnification from galaxy number counting is a challenging task. Meanwhile, the prominent non-Gaussian characteristic in late universe matter distribution disables the methodologies like the quadratic estimator \cite{Hu_2002, Maniyar_2021}. 
One of the possible solutions for galaxy clustering mitigation in magnification reconstruction is utilizing the magnification response in various flux bands \cite{zhang2005mapping}. This approach has been implemented to seek for the exact solution of lensing power spectrum \cite{Yang_2017, Zhang_2018} and the magnification estimation with minimum noise impact \cite{Yang_2011, Yang_2014, hou2021weak, ma2024method, qin2023weak}. 

\citet{zhang2019abs} proposed an analytic method for blind separation (ABS) to extract the CMB $B$-mode signal from blind foreground utilizing only the frequency dependence. In the context of CMB foreground removal, its robustness and accuracy were validated in the simulated sky maps, performing unbiased recovery of 
power spectrums \cite{Yao_2018, santos2021reconstructing}, and recently it is employed in the AliCPT-1 pipline for the primordial $B$-mode detection \cite{ghosh2022performance, zhang2024forecast}. Within the ABS framework, \citet{Yang_2017} and \citet{Zhang_2018} demonstrate that the cosmic magnification is strictly degenerated with galaxy clustering in solving the magnification power spectrum $C^{\kappa}_\ell$, but there is still a unique mathematical solution of $(1-r^2_{m\kappa})\,C^{\kappa}_\ell \;$, where $r_{m\kappa}$ is the cross correlation coefficient of matter and lensing convergence. 
However, as first noticed in this paper, their derivation neglects the intrinsic difference between galaxy bias and galaxy-lensing cross-correlation bias due to non-zero redshift width of the realistic photometric objects. Given that we aim to extract a tiny signal of cosmic magnification \cite{ziour2008magnification, duncan2022cosmological, elvin2023dark, euclidcollaboration2023euclid} from overwhelming galaxy intrinsic clustering, this overlooked difference invalidates ABS as the theoretically exact solution, and leads to non-negligible bias in practical application. This hidden systematic is a major point to convey in this paper. 

Alternatively, the lensing reconstruction can be realized by minimizing the contamination in estimation instead of exactly recovering. The constrained Internal Linear Combination (cILC, \cite{tegmark2003high, Remazeilles_2010, Remazeilles_2021}) is designed to eliminate the selected contamination and minimize the variance of residuals. It is semi-blind with certain prior knowledge of the foreground property, and the technique is proved effective for the mitigation of intrinsic galaxy clustering \cite{Yang_2011, Yang_2014,hou2021weak,qin2023weak, ma2024method}. In the context of magnification reconstruction, \citet{hou2021weak} proposed to reconstruct the weak lensing from galaxy overdensity through eliminating the mean galaxy bias, and subsequently the algorithm is implemented in DECaLS catalog \cite{qin2023weak}. \citet{ma2024method} updated the methodology by utilizing a scaling relation confirmed in \citet{zhou2023principal} to suppress the intrinsic galaxy clustering across the sub-samples. The improved algorithm reconstructs the weak lensing with significantly reduced intrinsic clustering, but the result is dominated by the shot noise in the sub-degree scale, and there is still $\sim 10$ percents bias in the cross-correlation with external field covering the redshift of source galaxies. 

In this work, we reveal the systematic in the current reconstruction algorithms, and attempt to demonstrate the challenges in extracting weak lensing from galaxy counting. In Section~\ref{sec:methodology}, we illustrate the intrinsic difference between the galaxy bias and galaxy-lensing cross-correlation bias, and demonstrate the risky approximation in disentangling the weak lensing signal for sub-division of galaxy samples. In Section~\ref{sec:ABSmethod}, we demonstrate the invalidity of ABS method in cosmic magnification reconstruction, as a result of the galaxy bias difference in photometric samples. In Section~\ref{sec:cILCmethod}, we update the algorithm proposed in Ref.~\cite{ma2024method}, and based on the simulation, we present its improvement in suppressing the galaxy clustering and reducing the shot noise. 

In the main text, our fiducial performance tests are based on the simulated cosmoDC2 galaxies \cite{korytov2019cosmodc2} within photometric redshift $0.8<z_{\rm ph}<1.2$ with surface number density $\bar{n}_g= 11.8\,{\rm arcmin}^{-2}$. We utilize their 6 flux bands $ugrizY$ and 15 galaxy color bands ($g-r,\,r-i,\cdots$) to sub-divide the galaxy samples, where there are 6 bins for a galaxy property and therefore total 126 galaxy sub-sample bins. The simulation details are listed in Appendix~\ref{sec:appendix_samples}. 
\textcolor{black}{In order to make our conclusions robust, we also present the re-analysis based on TNG300-1 galaxies \cite{Springel_2017, Nelson_2017, Pillepich_2017, Naiman_2018, Marinacci_2018} in Appendix~\ref{sec:tng300-results}, where we utilize different simulated galaxies but obtain similar results. }


\begin{figure*}
\includegraphics[width=1.0\textwidth]{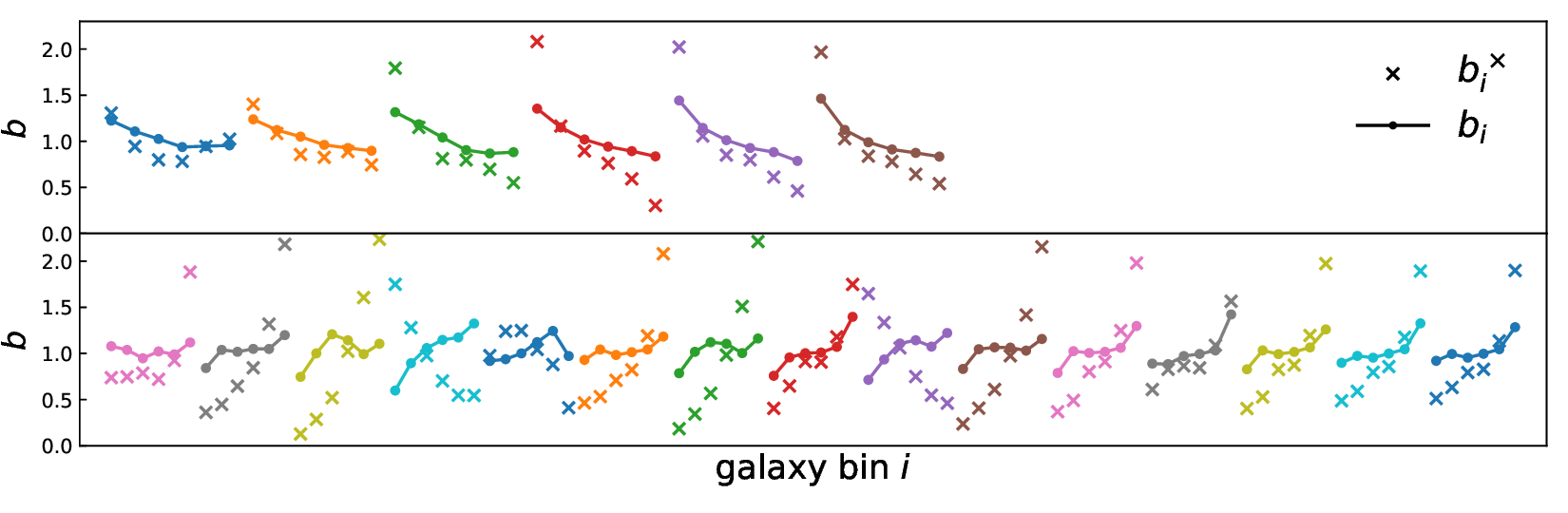}
\caption{ \label{fig:galaybias}
Galaxy bias $b_i \equiv C^{i m}_\ell /C^{m}_\ell$ (dot lines) and galaxy-lensing cross-correlation bias $b_i^\times \equiv C^{i \kappa}_\ell /C^{m\kappa}_\ell$ (cross markers) measured in on the fiducial galaxy samples at $\ell=700$. 
The upper panels show the sub-samples in 6 flux bands $ugrizY$, and the bottom panels show the sub-samples in 15 galaxy color bands, where there are $6$ bins for a given galaxy flux/color band. 
There are intrinsic difference between $b_i$ and $b^\times$, caused by (a) the mismatch between lensing kernel and the galaxy distribution; (b) the galaxy clustering evolution within the redshift bin. The difference between bias violates the hypothesis of ABS method. 
}
\end{figure*}

\section{Principle of Reconstruction and Its Systematic}
\label{sec:methodology}

Cosmic magnification in galaxy lensing modifies the observed flux and distorts the spatial galaxy number distribution. In the weak lensing region, the observed projected galaxy overdensity in linear order is \cite{Schneider_2006}
\begin{eqnarray}
    \label{equ:deltagL}
    \Delta_g^L = \Delta_g + q \kappa  \quad,
\end{eqnarray}
where $\Delta_g=\int{n_g(z)\delta_g(z)dz}$ is the intrinsic surface galaxy overdensity, $n_g(z)$ is the galaxy redshift distribution, and $\kappa=\int{W_\kappa(\chi)\delta_m(z)d\chi} \,$ is the weak lensing convergence. $q\equiv 2(\alpha-1)$ is the magnification response. For the flux limited samples, $\alpha$ is the logarithmic slope of galaxy luminosity function.
In order to disentangle $\kappa$, we divide lensed galaxy samples into various bins according to the galaxy properties, then the cross power spectrum of $i$-th and $j$-th galaxy sub-sample bins is, 
\begin{eqnarray}
    C^{ij, L}_\ell = C^{ij}_\ell + q_i C^{j\kappa}_\ell + q_j C^{i\kappa}_\ell + q_iq_j C^{\kappa}_\ell   \quad.
\end{eqnarray}
Throughout this paper, the indices $i,\,j$ correspond to $i,\,j$-th galaxy sub-sample bin, and $\ell$ is multipole. We can decompose the intrinsic galaxy clustering into $\Delta_{g,i} = b_i\Delta_m + \Delta_{i,\mathcal{S}}$, where $\Delta_m$ is the underlying matter clustering and $\Delta_{i,\mathcal{S}}$ is the stochastic component in the galaxy clustering with $\langle \Delta_{i,\mathcal{S}} \Delta_{m}^* \rangle=0$. Therefore 
\begin{eqnarray}
    \label{equ:clgg_detm}
    C^{ij, L}_\ell = b_i b_j C^{m}_\ell + (q_ib_j^\times + q_jb_i^\times) C^{m\kappa}_\ell+ q_iq_j C^{\kappa}_\ell + C_\ell^{ij,\mathcal{S}} 
\end{eqnarray}
The galaxy deterministic bias is $b_i \equiv C^{i m}_\ell /C^{m}_\ell$ and the galaxy-lensing cross-correlation bias is $b_i^\times \equiv C^{i \kappa}_\ell /C^{m\kappa}_\ell$. The stochastic power spectrum $C_\ell^{ij,\mathcal{S}}$ is blind to us due to the complex halo collapse, galaxy formation and the non-Poisson nature of galaxy \cite{Pen_2003, seljak2009suppress, bonoli2009halo, Hamaus_2010, desjacques2018large}. 

If the source galaxies distribute along the redshift with a $\delta$-function extent, two biases $b_i^\times$ and $b_i$ is exactly same. But in order to suppress the fluctuation from the discrete sampling noise, sufficient counting quantity is required and the corresponding galaxy distribution is spread along redshift. The galaxy bias $b_i$ is the integration over $z$ weighted by $n_g(z)$ while the bias $b_i^\times$ is weighted by lensing kernel $W_\kappa(\chi)$. The former intrinsic galaxy probes more the contribution around $n_g(z)$ peak while the latter probes lower redshift contribution \cite{wenzl2023magnification}, therefore $b_i$ and $b_i^\times$ are generally different due to the galaxy evolution. The distinguishing redshift evolutions among different galaxy sub-samples result in various deviation behaviors between $b_i$ and $b_i^\times$, as shown in Fig.~\ref{fig:galaybias}. This substantial distinction is neglected in previous reconstruction algorithm though magnification response, and it is a non-negligible systematic that should be considered. Additionally, similar problems due to the mismatching kernels have also been reported in other statistics (e.g. $E_G$ statistic \cite{Zhang_2007, wenzl2024constraining} and photometric error correction \cite{saraf2024effect}), but it introduces a worse situation in our case because of the mixture of auto and cross signals.

Another potential risk is that the non-negligible redshift evolution also results in the distinguishing galaxy redshift distribution $n_{g}(z)$ in each bin. The discrimination is minor for the flux divided sub-samples, but significant for the galaxy color divided sub-samples (refer to Appendix.~\ref{sec:appendix_samples_2} for details). It partially explains the severe distinction between $b_i$ and $b_i^\times$ for color divided sub-samples in Fig.~\ref{fig:galaybias}, that different $n_{g}(z)$ profiles lead to redshift weights in the projecting integration. Specifically, if the galaxy distribution skews to the lower redshift then the matter-lensing matching efficiency increases, thus the cross-correlation signal is enhanced and the measured bias is larger, vice versa. On the other hand, different $n_{g}(z)$ profile generates different magnification responses, that the lensing signals in different bins are potentially different, that $\kappa_i\neq\kappa_j$ if $i\neq j$. It risks the fundamental principle of reconstruction that lensing signal $\kappa$ is the same in different sub-samples. Fortunately, it is not an essential problem, since with prior knowledge of galaxy evolution and their redshift distribution, we can apply appropriate redshift cut as well as redshift weight to match the redshift distribution of each sub-sample coercively. In principle, it would be solved with future more accurate photometric redshift. Therefore we avoid this complexity and still treat lensing signal $\kappa$ for different sub-sample bins as same, and utilize all the galaxy magnitude and color information, intending to provide optimistic results. But we caution that it should be carefully processed in the realistic application.


\begin{figure*}
\includegraphics[width=0.9\textwidth]{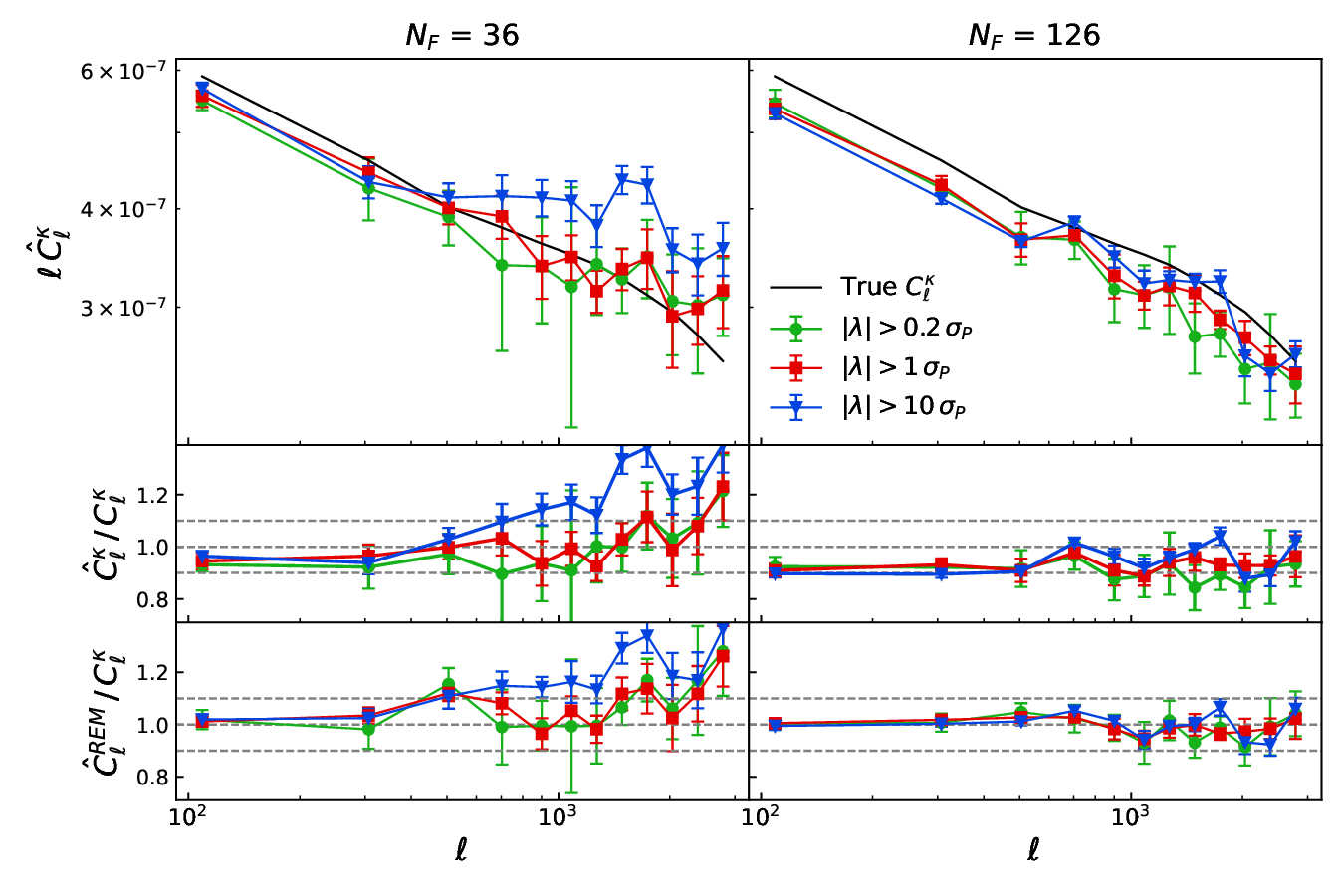}
\caption{ \label{fig:recont_ABS} 
ABS reconstruction performance on fiducial galaxy samples. 
The left panels show the results reconstructed by 6 flux bands (total $36$ bins), and the right panels show the results reconstructed 6 flux $\&$ 15 galaxy color bands (total $216$ bins), where there are $6$ bins for a given galaxy flux/color band. 
The upper panels compare the reconstruction using eigenvalue threshold $\lambda_{\rm cut}=0.2/1/10\;\sigma_P$ with true $C^\kappa_\ell$ signal, and the middle panels show their ratios. The bottom panels show the results by artificially removing $C^{\kappa g}_\ell$ contribution from $C_\ell^{ij,L}$.
The dash lines in lower panels denote $\pm 10\%$ region. 
}
\end{figure*}

\section{Reconstruction Algorithm: ABS}
\label{sec:ABSmethod}

ABS method \cite{zhang2019abs} aims to separate the signal from the unknown foreground using only the spectral information. It states that in each multipole $\ell$ bin, 
\begin{eqnarray}
    \label{equ:abs_equ}
    {\mathcal D}_{ij}(\ell) =f_i f_j {\mathcal D}_B(\ell) + {\mathcal D}_{ij}^{\rm fore}(\ell)  \;,
\end{eqnarray}
where $f_i$ is the frequency dependence in $i$-band, $i=1,\cdots,N$, the only known information. The cross band power matrix ${\mathcal D}_{ij}$ is the observable, and ${\mathcal D}_B$ is the signal aimed to reconstruct. ${\mathcal D}_{ij}^{\rm fore}(\ell)$ is the unknown foreground matrix. If 
\begin{itemize}
\item[\bf{(i)}]  The frequency dependence vector $f_i$ is not contained in the subspace spanned by the eigenvectors of the foreground matrix ${\mathcal D}_{ij}^{\rm fore}$ ;
\item[\bf{(ii)}] The rank $M$ of the foreground matrix ${\mathcal D}_{ij}^{\rm fore}$ is smaller than the size $N$, $M<N$ ;
\end{itemize}
the equations Eq.~(\ref{equ:abs_equ}) are solvable, and the solution of ${\mathcal D}_B$ is unique, given by
\begin{eqnarray}
    \label{equ:abs_sol}
    {\mathcal D}_B = \left( \sum_{\mu=1}^{M+1} G_\mu^2 \lambda_{\mu}^{-1} \right)^{-1}   \quad .
\end{eqnarray}
Here we make an eigenvalue decomposition, ${\mathcal D}_{ij} = \sum_{\mu}{ \lambda_{\mu} E^{(\mu)}_i E^{(\mu)}_j }$, where $\lambda_{\mu}$ is the $\mu$-th eigenvalue and $E^{(\mu)}_i$ is the $i$-th element of the $\mu$-th eigenvector, and $G_\mu = \sum_i{ f_i E^{(\mu)}_i }$. Notice that two hypothesies infer that the rank of $\mathcal{D}_{ij}$ is $M+1$, thus the summation in Eq.~(\ref{equ:abs_sol}) ranges from $1$ to $M+1$.

In order to exclude the unphysical modes and the small eigenvalues dominated by noise, we can shift the cross band power by ${\mathcal D}_{ij} \;\rightarrow\; {\mathcal D}_{ij} + f_i f_j\mathcal{S}$ and the estimation is modified as  \cite{zhang2019abs, Zhang_2018},
\begin{eqnarray}
    \label{equ:abs_sol_S}
   \hat{\mathcal D}_B = \left( \sum_{\lambda_\mu > \lambda_{\rm cut}} G_\mu^2 \lambda_{\mu}^{-1} \right)^{-1}_{\rm shifted} - \mathcal{S}  \quad,
\end{eqnarray}
where the shifted parameter $\mathcal{S}$ and eigenvalue threshold $\lambda_{\rm cut}$ are artificially chosen to stabilize the estimation.

Because there are only several effective eigen-components in the galaxy stochasticity \cite{bonoli2009halo, Hamaus_2010, zhou2023principal} and the vector $q_i$ is uncorrelated with the eigenmodes of the matrix $C_\ell^{ij,\mathcal{S}}$, ABS method naturally removes the stochasticity contribution $C_\ell^{ij,\mathcal{S}}$ with sufficient galaxy sub-sample bins. Thus in the following derivation, we can neglect the $C_\ell^{ij,\mathcal{S}}$ term, and the remained unknown variables are $\left\{  C^\kappa_\ell \;,\; (C^{m}_\ell)^{1\over 2}\,b_i \;,\; C^{m\kappa}_\ell\,b_i^\times \right\}$. For illustration, we recombine the terms in Eq.~(\ref{equ:clgg_detm}) as 
\begin{eqnarray}
    \label{equ:clgg_detm_recomb}
    C^{ij, L}_\ell &=& q_iq_j C^{\kappa}_\ell \\
    \nonumber &+& C^{m}_\ell\, b_i b_j + {1\over2}C^{m\kappa}_\ell (q_i + b_i^\times)(q_j + b_j^\times)  \\
    \nonumber &-& {1\over2}C^{m\kappa}_\ell (q_i - b_i^\times)(q_j - b_j^\times)  \quad.
\end{eqnarray}
If we aim to reconstruct $C^\kappa_\ell$ with magnification response vector $q_i$ as known, we recognize the second and third lines of Eq.~(\ref{equ:clgg_detm_recomb}) is the foreground matrix in terms of ABS. Its eigenspace associated with non-zeros eigenvalues are spanned by vector ${\bf b}$, ${\bf q} +{\bf b}^\times$ and ${\bf q} - {\bf b}^\times$, and obviously the vector ${\bf q}$ is contained in the subspace spanned by these three vectors, violating the hypothesis {\bf (i)}. The conclusion remains even though $b_i=b_i^\times$, where now vector ${\bf b}$ and ${\bf q}$ span the subspace of foreground eigenvectors, and it recovers the strict variable degeneracy in Ref.~\cite{Yang_2014}. Therefore, we confirm once again that it is impossible to separate $C^\kappa_\ell$ blindly with only frequency dependence.

A possible solution for reducing degeneracy is to perform variable substitution \cite{Yang_2014}. We adopt the same replacement as Ref.~\cite{Yang_2014}, 
\begin{eqnarray}
    \nonumber&& \tilde{b}_i = b_i + q_i{ C^{m\kappa}_\ell\over C^m_\ell} \;,\quad  \tilde{b}_i^\times = b_i^\times + q_i{ C^{m\kappa}_\ell\over C^m_\ell}   \;,   \\
    \nonumber&& \tilde{C}^\kappa_\ell = C^\kappa (1-r^2_{m\kappa} )   \;,\quad  r^2_{m\kappa} = { \left(C^{m\kappa}_\ell\right)^2 \over C^{m}_\ell C^\kappa_\ell }  \;,
\end{eqnarray}
and we obtain 
\begin{eqnarray}
    \label{equ:clgg_detm_recomb2}
    C^{ij, L}_\ell &=& q_i q_j \tilde{C}^\kappa_\ell  \\
    \nonumber &+& C^{m}_\ell\, \tilde{b}_i \tilde{b}_j  + C^{m\kappa}_\ell \left[ (\tilde{b}_i^\times -\tilde{b}_i)q_j + (\tilde{b}_j^\times -\tilde{b}_j)q_i \right]    \quad.
\end{eqnarray}
If we aim to reconstruct $\tilde{C}^\kappa_\ell$, the unknown variables are $\left\{  \tilde{C}^\kappa_\ell \;,\; (C^{m}_\ell)^{1\over 2}\,\tilde{b}_i \;,\; C^{m\kappa}_\ell \left(\tilde{b}_i^\times -\tilde{b}_i\right) \right\}$. Similarly, we recognize that vector ${\bf q}$, $\tilde{\bf b}$ and ${\bf b}^\times -{\bf b}$ span the foreground eigenspace associated with non-zeros eigenvalue, supposing that these vectors are not degenerate. Unfortunately, it still violates hypothesis {\bf (i)}, and declares that it is impossible to solve $\tilde{C}^\kappa_\ell$ in principle. Further techniques no longer work for reducing the variables degeneracy, since the redundant terms can not be grouped into rank one matrix. If we neglect the difference between two bias, approximately $b_i^\times = b_i$, the only non-vanishing component of foreground matrix is $(C^{m}_\ell)^{1\over 2}\,\tilde{b}_i$. In this case, as adopted in Ref.~\cite{Yang_2014, Yang_2017}, we can solve for $\tilde{C}^\kappa_\ell$ from Eq.~(\ref{equ:clgg_detm_recomb2}).

The degeneracy in the cosmic magnification separation is caused by the cross-correlation term $C^{m\kappa}_\ell$ in Eq.~(\ref{equ:clgg_detm_recomb2}). Any difference appearing in $b_i$ and $b_i^\times$ would mix the $q_i$ vector with the foreground matrix, then results in the inseparation. However, in the realistic measurement, the noise distorts the cross power matrix and changes the distribution of the components consisting of the observable, and the cross-correlation mixture contribution is minor compared with prominent intrinsic galaxy clustering. Therefore in the present of noise, we expect the $q_i$ dependence would be misaligned in foreground components. In this perspective, the mathematical solutions Eq.~(\ref{equ:abs_sol}) $\&$ (\ref{equ:abs_sol_S}) are not longer rigorous, but still serve as a rough estimation for cosmic magnification power spectrum to some extent.

We show the performance test of ABS reconstruction in Fig.~\ref{fig:recont_ABS}. We subtract the shot noise expectation before applying the ABS reconstruction, and set the shift parameter as $\mathcal{S}=20\sigma_P$ to reduce the impact of residual noise fluctuation, where we define $ \sigma_P = \left((2\ell+1)\Delta\ell f_{\rm sky}/2\right)^{-1/2} C_N$ and $C_N$ is the shot noise expectation in each bin \cite{zhang2019abs,Yang_2017,Zhang_2018}. We filter the noise-dominating eigenmodes by selecting eigenmodes with $|\lambda|>\lambda_{cut}\,$, and the absolute value is taken because of the negative eigenmodes due to the sub-Poisson behavior in stochasticity \cite{Hamaus_2010}. As shown in Fig.~\ref{fig:recont_ABS}, the reconstruction using only the flux information is dominated by shot noise, and is sensitive to the choice of $\lambda_{\rm cut}$. When the eigenvalue threshold increases, the statistic error is reduced but the systematic error is high. The large scale reconstructions (the first 2 $\ell$ bins) are relatively stable \textcolor{black}{because of signal domination}, but there is about $5\sim 10 \%$ systematic under-estimation. While for the reconstruction using both the flux and color information, the reconstruction is robust alongside the eigenvalue threshold, and the statistic error is significantly reduced. It is a demonstration that more degree of freedom in a set of equations, the solution is more robust. Though the statistic fluctuation is suppressed, the systematic is striking as $10\%$ low, exceeding that from $r^2_{m\kappa}\lesssim 5\%$. 
\textcolor{black}{Such an underestimation is expected, since the majority of the magnification responses $q_i$ are negative values, equivalent to a negative degeneracy between the lensing signal and galaxy clustering. Therefore, the lensing signal is leaked to the foreground matrix to compensate the negative degeneracy, and it leads to the systematic underestimation in estimates.}
The results confirm our inference that the degeneracy due to the galaxy-lensing correlation breaks down the ABS exact solution, and the noise in observable distorts the eigenmodes distribution that ABS still serves as rough but biased estimation of lensing power spectrum.

In Fig.~\ref{fig:recont_ABS}, we also present the reconstruction $\hat{C}^{\rm REM}_\ell$ by artificially removing the $C^{\kappa g}_\ell$ contribution, that $ \hat{C}^{ij,L}_\ell \,\rightarrow\, \hat{C}^{ij,L}_\ell - q_i C^{j\kappa}_\ell -q_j C^{i\kappa}_\ell$. In this case, we expect the ABS solution is exact if the number of galaxy bins is larger than the number of galaxy stochasticity eigenmodes, since we remove the degenerate components. The figure shows that both the reconstruction with flux-only or flux$\,\&\,$color can recover an unbiased lensing power spectrum \textcolor{black}{in large scale, where} the remarkable $10\%$ systematic underestimation is removed.
\textcolor{black}{In the small scale with $\lambda_{\rm cut}\leq 1\sigma_P$, the results with flux-only are unstable and biased, while those with flux$\,\&\,$color is able to perform almost unbiased recovery within $100\lesssim\ell\lesssim 3000$.}
We suppose the effective degree of freedom in galaxy stochasticity is approximately several tens due to projected $k$ modes mixture, which can not be \textcolor{black}{robustly} separated by $36$ sub-sample bins but resolved with $126$ bins. It verifies the conclusion that it is the galaxy-lensing degeneracy that prevents us from disentangling cosmic magnification, but not the stochasticity, where the latter can be solved with enough independent bins.

\textcolor{black}{Switching the eigenvalue threshold to $\lambda_{\rm cut}=10\sigma_P$ in the reconstruction with flux$\,\&\,$color, there is a noticeable oscillating feature at $\ell\gtrsim 1000$. This is sourced by the improper large $\lambda_{\rm cut}$, which filters not only more noise-dominating eigenmodes but also partial lensing signal. Specifically, shot noise is sub-dominant in large scale thus has negligible impact on reconstruction. While in small scale, shot noise overwhelms the small eigenmodes which contain the lensing signal, therefore the discard of these noise-dominating modes leads to the systematic bias \cite{zhang2019abs,Yang_2017,Zhang_2018}. Meanwhile, the positive and negative eigenvalues alternating around the threshold leads to an oscillating trend. To mitigate the loss of signal, we can adopt a conservative choice of threshold, i.e., $\lambda_{\rm cut}\lesssim 1\sigma_P$ for fiducial samples. Similar problem has a more serious impact on the reconstruction using TNG300-1 samples, where the shot noise dominates the reconstruction performance at small scale. Specifically, there is fewer effective degree of freedom in the TNG300-1 samples, thus the eigenvalue distribution is steep and partial signal is contained in tiny eigenmodes. In the small scale, the threshold $\lambda_{\rm cut}\sim 1\sigma_P$ excludes the tiny eigenvalues to avoid the noise sourcing reconstruction divergence, while the signals in these tiny eigenmodes are no longer recovered, resulting in the overestimation. 
It also explains why, for the flux-only reconstruction at $\ell\gtrsim1000$ in Fig.~\ref{fig:recont_ABS}, the result with $\lambda_{\rm cut}=10\sigma_P$ is biased with reduced uncertainty compared to that with $\lambda_{\rm cut}=1\sigma_P$. 
}


\begin{figure*}
\includegraphics[width=0.8\textwidth]{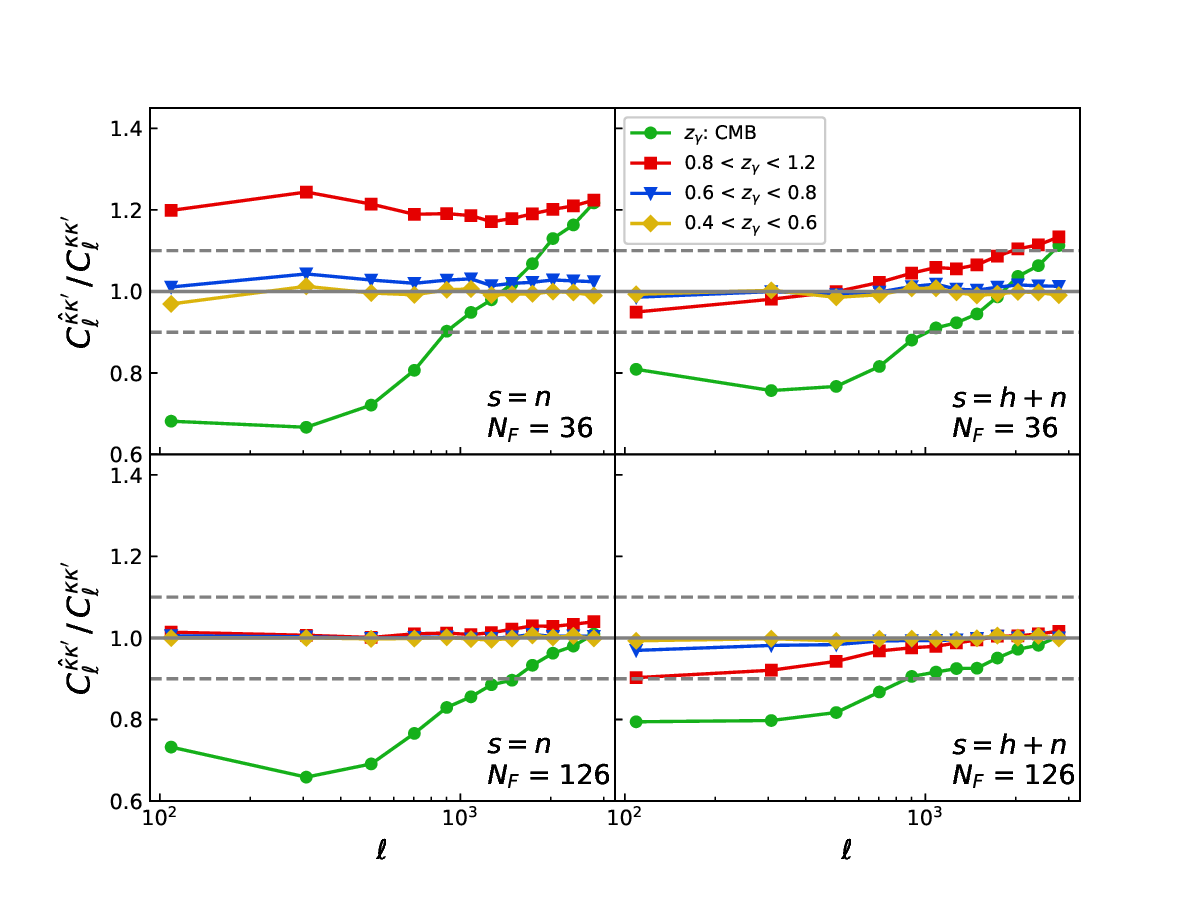}
\caption{ \label{fig:cILC_cross} 
cILC reconstruction performance on fiducial galaxy samples. We present the cross correlation results between the reconstructed $\hat{\kappa}$ and the external $\kappa^\prime$ map from different redshift source. 
The upper panels show the results reconstructed by 6 flux bands (total $36$ bins), and the bottom panels show the results reconstructed 6 flux $\&$ 15 galaxy color bands (total $216$ bins), where there are $6$ bins for a given galaxy flux/color band. 
The left panels show the results with label $s=n$ using Eq.~(\ref{equ:cILC_condi2}), and the right panels show the results with label $s=h+n$ using Eq.~(\ref{equ:cILC_condi1}).
The dash lines in lower panels denote $\pm 10\%$ region. 
}
\end{figure*}
\begin{figure}
\includegraphics[width=1.0\columnwidth]{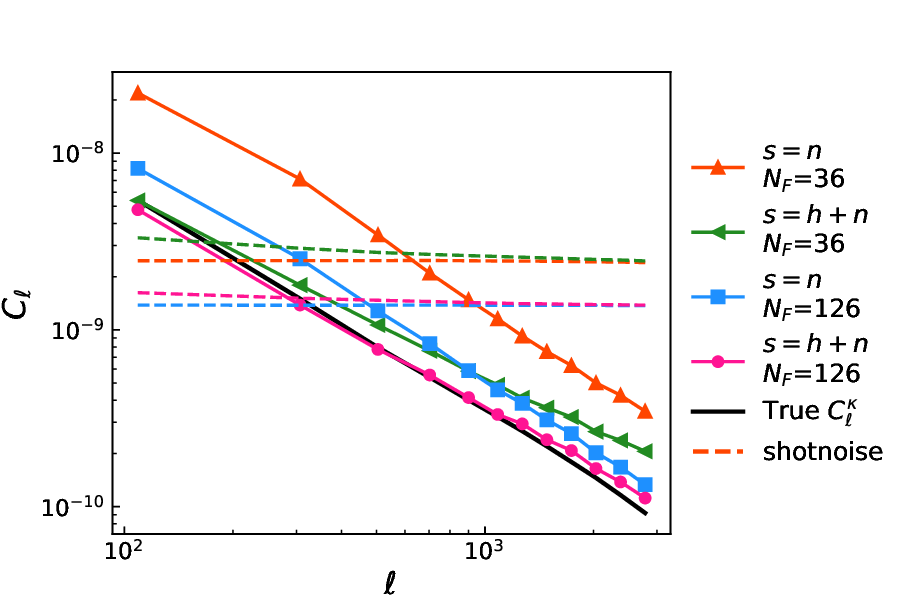}
\caption{ \label{fig:cILC_auto} 
cILC reconstruction performance on fiducial galaxy samples. We present the reconstructed cosmic magnification power spectrum $\hat{C}^{\kappa}_\ell \equiv C^{\hat{\kappa}}_\ell - C^N_\ell$ (solid lines) and the corresponding shot noise $C^N_\ell$ (dash lines), where we adopt the same labels as in Fig.~\ref{fig:cILC_cross} for different schemes.
}
\end{figure}

\section{Reconstruction Algorithm: cILC}
\label{sec:cILCmethod}

The cILC method aims to reconstruct the signal by superimposing different frequent maps with appropriate weights, requiring eliminating the selected contaminating components while minimizing the noise. In the spherical harmonic space, we construct the estimator 
\begin{eqnarray}
\hat{\kappa}_\ell &=& \sum_{i} w_i(\ell) \Delta_{g, i}^L(\ell)  \quad,
\end{eqnarray}
and the corresponding power spectrum 
\begin{eqnarray}
&& C^{\hat{\kappa}\hat{\kappa}}_\ell = \left(\sum_{ij} w_i q_i\right)^2 C^{\kappa\kappa}_\ell \\
&& \nonumber +\, \sum_{ij} w_i w_j \left(q_j C^{i\kappa}_\ell +q_i C^{j\kappa}_\ell \right)  + \sum_{ij} w_i w_j \left( C^{ij}_\ell + {\bar{n}_{ij}\over \bar{n}_i \bar{n}_j } \right) 
\end{eqnarray}
where we have explicitly written down the contribution from shot noise $\,{\bar{n}_{ij} /\left(\bar{n}_i \bar{n}_j\right)}$ for $i,j$-band power spectrum. We can decompose the observed cross band power as 
\begin{eqnarray}
\hat{C}^{ij,L}_\ell = \lambda_{1} E^{(1)}_i E^{(1)}_j + \sum_{\mu>1}{ \lambda_{\mu} E^{(\mu)}_i E^{(\mu)}_j } + {\bar{n}_{ij}\over \bar{n}_i \bar{n}_j }   \quad,
\end{eqnarray}
here the ensemble shot noise has been subtracted before eigen-decomposition. Ref.~\cite{zhou2023principal} have confirmed the proportional relation between galaxy deterministic bias and the first eigenvectors, 
\begin{eqnarray}
b_i \propto E^{(1)}_i   \quad,
\end{eqnarray}
and other higher order eigen-components are dominated by the galaxy stochasticity. We further postulate that the galaxy-lensing cross correlation bias is equal to galaxy deterministic bias. We can reconstruct the ${\kappa}$ by following constraints on the weight $w_i$ \cite{ma2024method}.
\begin{eqnarray}
\label{equ:constrains1} && \sum_i w_i q_i = 1 \\
\label{equ:constrains2} && \sum_i w_i E^{(1)}_i = 0   \\
\label{equ:constrains3} && {\rm minimize\;} N = \sum_{ij} w_i w_j s_{ij}
\end{eqnarray}
The first condition aims to eliminate the multiplicative bias, and the second condition aims to eliminate the intrinsic galaxy clustering. The last condition intends to minimize all other components, including the galaxy stochasticity and shot noise, with
\begin{equation}
    \label{equ:cILC_condi1}
    s_{ij} = \sum_{\mu>1}{ \lambda_{\mu} E^{(\mu)}_i E^{(\mu)}_j } + {\bar{n}_{ij}\over \bar{n}_i \bar{n}_j }   \quad .
\end{equation}
Conservatively, we can choose to minimize only the shot noise contribution, with
\begin{eqnarray}
    \label{equ:cILC_condi2}
    s_{ij} = {\bar{n}_{ij}\over \bar{n}_i \bar{n}_j }  \quad,
\end{eqnarray}
as the scheme adopted by Ref.~\cite{ma2024method}. The general solution is given by ${\bf w = e^T (A^T N^{-1}A)^{-1} A^T N^{-1}}$, where ${\bf e^T\equiv}[\,1\,0\,]$ and matrix ${\bf A\equiv\left[\,q\,E\,\right]}$ is stacked by column $q_i$ and column $E_i^{(1)}$ in our case \cite{Remazeilles_2010, Remazeilles_2021, ma2024method}.

In Fig.~\ref{fig:cILC_cross}, we present the reconstruction performance by cross correlating the reconstructed $\hat{\kappa}$ with cosmic shear map or CMB lensing map. We recover the similar systematic error behavior as that found in Ref.~\cite{ma2024method}, that is negative systematic bias in large scale while positive systematic in small scale. The large scale bias could be caused by the chance correlation inherited in the ILC method, or induced by the constraint condition Eq.~(\ref{equ:constrains2}) of eliminating the first principal component of the cross power spectrum. Specifically, though the eigen-decomposition projects almost all the intrinsic clustering signal into the first eigenmode so that we can alleviate galaxy clustering contamination by eliminating $E^{(1)}_i$, it also 
leaks some lensing signal into the $E^{(1)}_i$ eigenmode. Thus enforcing Eq.~(\ref{equ:constrains2}) to eliminate $E^{(1)}_i$ may leads to the loss of lensing signal, and results in the observed underestimation of the large scale cross-correlation with external field $z_\gamma\gtrsim z_\kappa$, as shown in Fig.~\ref{fig:cILC_cross}.   
Besides, because the matter distribution is correlated with the cosmic shear from higher redshift source, there is significant overestimation in the small scale and in the reconstruction scheme with Eq.~(\ref{equ:cILC_condi1}) $\&$ $N_F=36$, due to the residual clustering in the reconstructed map. 
\textcolor{black}{While for the cross-correlation with cosmic shear in lower redshift, the correlated part in the reconstructed lensing map is fixed by constraint condition Eq.~(\ref{equ:constrains1}) and the galaxy clustering contamination is not correlated with the cosmic shear in lower redshift, thus the cross-correlation is almost unbiased for all four cases in Fig.~\ref{fig:cILC_cross}.}
It is remarkable that both the inclusion of more galaxy bins or higher eigen-components in Eq.~(\ref{equ:cILC_condi1}) can suppress the intrinsic clustering significantly. However, there is a danger that the suppression of intrinsic clustering also leaks the lensing signal around $z_\kappa$, and the hidden systematic is ambiguous to quantify due to the lack of knowledge about galaxy stochasticity in reality.

In Fig.~\ref{fig:cILC_auto}, we present the reconstructed $\hat{\kappa}$ auto power spectrum $\hat{C}^{\kappa}_\ell \equiv C^{\hat{\kappa}}_\ell -C^N_\ell$ with shot noise subtraction and the corresponding shot noise $C^N_\ell$. The reconstruction using only the flux information with Eq.~(\ref{equ:cILC_condi2}) is significantly overestimated, due to the residual clustering contamination. Both the inclusion of more galaxy bins or higher eigen-components can alleviate the overestimation, but still be significantly biased. The reconstruction using both flux$\,\&\,$color information with Eq.~(\ref{equ:cILC_condi1}) can realize promising suppression of the residual noise, but it underestimates the large scale power spectrum due to the leak of lensing power. Besides, the reconstruction with more independent galaxy bins can significantly suppress the shot noise, where the shot noise dominating scales increase from $\ell\gtrsim 150$ to $\ell\gtrsim 300$. It is an illustration that the intrinsic discrete sampling noise can be reduced by separating the galaxy clustering \cite{seljak2009suppress}. 
\textcolor{black}{Furthermore, we replace the fiducial samples with the TNG300-1 samples, and the overall performance by cILC method are roughly the same as the fiducial one, in both the cross-correlation and auto-correlation results. The unbiased cross-correlation with cosmic shear in lower redshift and the underestimated cross-correlation with CMB lensing are recovered. But the details, such as the proportion of suppressing matter clustering, exhibit obvious distinction, and it also illustrates that the specific reconstruction performance depends on the galaxy samples.}


\section{Discussion and Conclusions}

In this work, we reveal a hidden systematic in ABS and cILC algorithms for reconstructing the cosmic magnification using the magnification response. 
We demonstrate that due to the mismatch of the weak lensing kernel and the galaxy distribution, the galaxy bias inferred from auto-correlation differs from the galaxy-lensing cross-correlation bias. It violates the basic principle of ABS method, leading to the mathematical inseparation of the cosmic magnification. Due to the presence of noise in the cross power matrix, the eigen-components are redistributed, and the degeneracy is alleviated to some extent. 
\textcolor{black}{However, the shot noise dominates the tiny eigenmodes and also causes the reconstruction instability in small scale.}
Based on simulation, we show that the ABS method allows a rough and biased estimation of comic magnification power spectrum. In principle, we can adopt narrow enough redshift bins to decrease the overlap of the lensing kernel and galaxy distribution, as well as to match the galaxy biases between galaxy auto-correlation and galaxy-lensing cross-correlation, so that ABS is recovered as an exact solution. But it is inevitable to decrease the number of galaxies and amplify the shot noise contamination, resulting into the overwhelming of noise fluctuation in the reconstruction. We emphasize that compared to realistic survey ability, we have adopted the extremely optimistic choice for the number density in the performance test, but it is still risk in excluding noise-dominating eigenmodes. 

We update the cLIC magnification reconstruction by including the color information and the higher order components to suppress the noise in reconstruction. We show that more independent galaxy bins from color information are able to suppress shot noise and intrinsic clustering, and the minimizing function including higher order eigen-components also reduces the residual matter clustering significantly. However, cILC method performs naturally biased recovering because of \textcolor{black}{the residual of matter clustering and} the leak of the lensing signal around the source redshift. It is obscure to blindly determine the fraction of biased power spectrum due to complex galaxy stochasticity. Analogously as the solution investigated in the CMB \cite{Saha_2008}, it is possible to de-bias cILC reconstructed results, but demanding further investigation in the future for the galaxy properties and the complex stochasticity evolution on large scale.
\textcolor{black}{Nevertheless, the contamination only impacts on the redshift range of the galaxy sources for reconstruction. Therefore, it is prospective to apply the cILC products, in which the shot noise is significantly suppressed, to obtain unbiased cross-correlation measurement with lower-redshift probes.}

In contrast to CMB lensing reconstruction, where the primary anisotropy is Gaussian field and the lensing signal is decorrelated with the primary fluctuation, the cosmic magnification field enters the observations coupled with the prominent non-Gaussian galaxy clustering, and all the discrete nature of galaxy, galaxy-lensing cross-correlation and non-linear evolution of matter clustering challenge the signal separation. 
\textcolor{black}{Any small residual contamination from theoretical deviation in the noise model may significantly bias the tiny cosmic magnification reconstruction.}
It remains highly challenging, even in theory, to reconstruct unbiased cosmic magnification signal solely from galaxy counting.
\textcolor{black}{One possible solution to mitigate issues in reconstruction is to apply prior knowledge from analytic theory or numerical simulation. For example, we can perform the reconstruction in the galaxy sample mocks to quantify the potential contamination and signal leakage, from which we can construct a transfer function to compensate for systematic bias in realistic data. Although this compensation scheme does not solve fundamental issues, it is practical for correcting the amplitude of the rough estimates at this stage.
Another possible remedy is to construct estimator combined with other cosmological probes or cosmological information. For example, with the aid of cosmic shear, we are able to infer galaxy-lensing cross-correlation bias and alleviate the systematic bias induced by galaxy bias difference, while the associated estimator and its applicability and performance require significant further investigation.
}


\begin{acknowledgments}
We would like to thank Yu Yu's help in the access and usage of simulations. 
This work is supported by  the National Key R\&D Program of China (2023YFA1607800,2023YFA1607801,2020YFC2201602), the National Science Foundation of China (11621303), CMS-CSST-2021-A02, and the Fundamental Research Funds for the Central Universities. 
This work made use of the Gravity Supercomputer at the Department of Astronomy, Shanghai Jiao Tong University. 

\end{acknowledgments}


\appendix

\begin{figure*}
\includegraphics[width=0.95\textwidth]{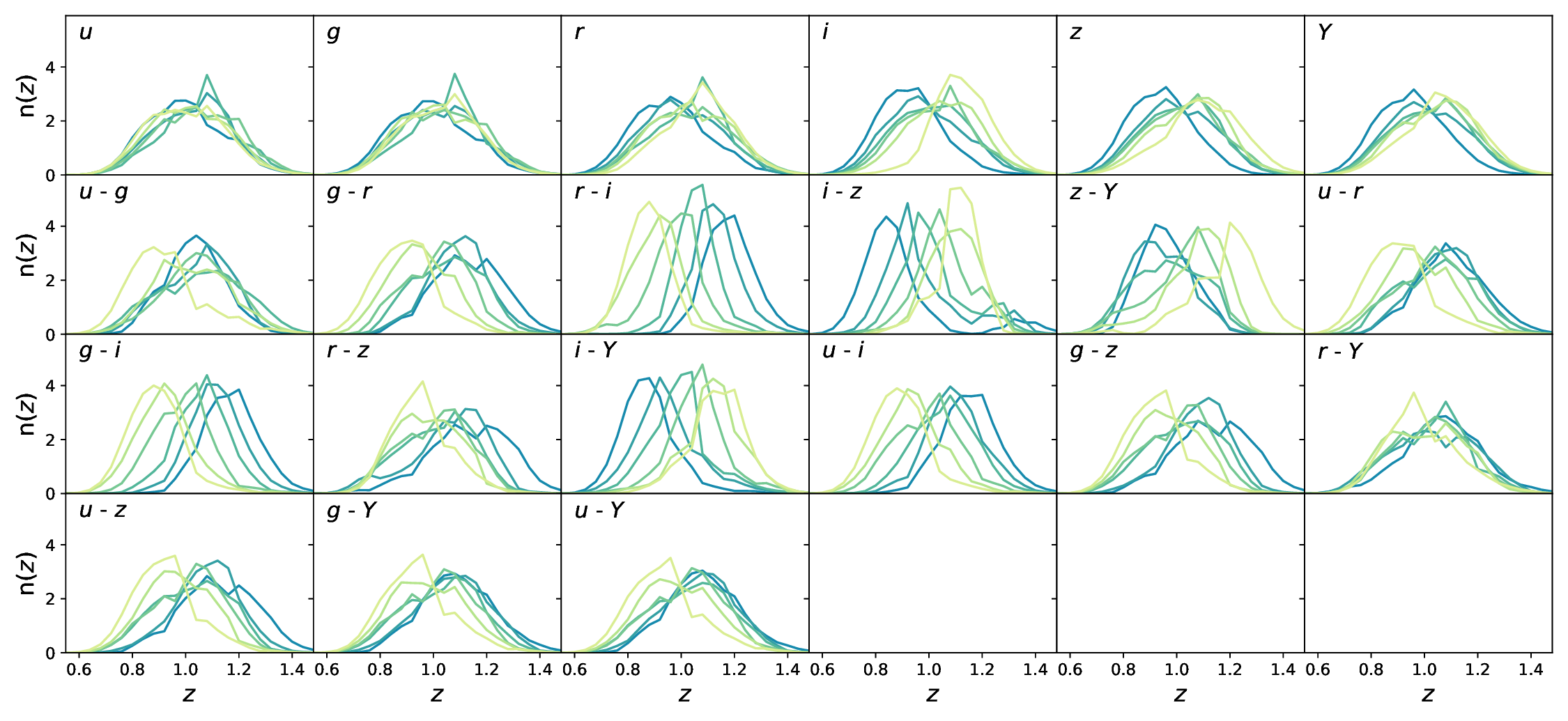}
\caption{ \label{fig:nz_cosmoDC2} 
The cosmoDC2 galaxy distribution $n_g(z)$ in photometric redshift range $0.8<z_{\rm ph}<1.2$ with scatter $\sigma_z=0.05(1+z)$. The galaxies are divided into 6 bins for given a galaxy property. The $n_g(z)$ profiles with more yellow in color denote the galaxy population with larger value of galaxy magnitude/color.
}
\end{figure*}
\begin{figure*}
\includegraphics[width=0.95\textwidth]{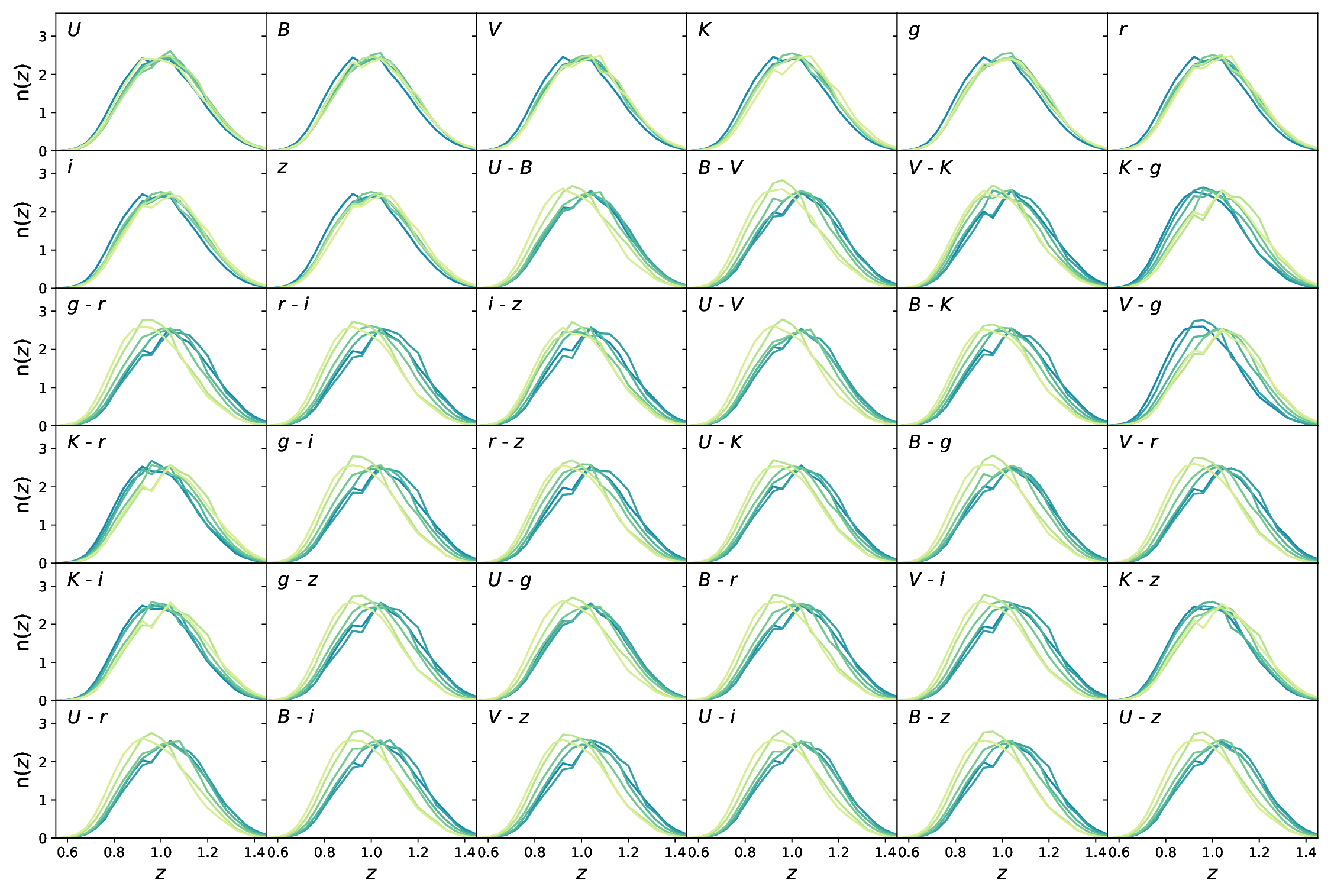}
\caption{ \label{fig:nz_tng300} 
The TNG300-1 galaxy distribution $n_g(z)$ in photometric redshift range $0.8<z_{\rm ph}<1.2$ with scatter $\sigma_z=0.05(1+z)$. Same as Fig.~\ref{fig:nz_cosmoDC2}.
}
\end{figure*}
\begin{figure*}
\includegraphics[width=1.0\textwidth]{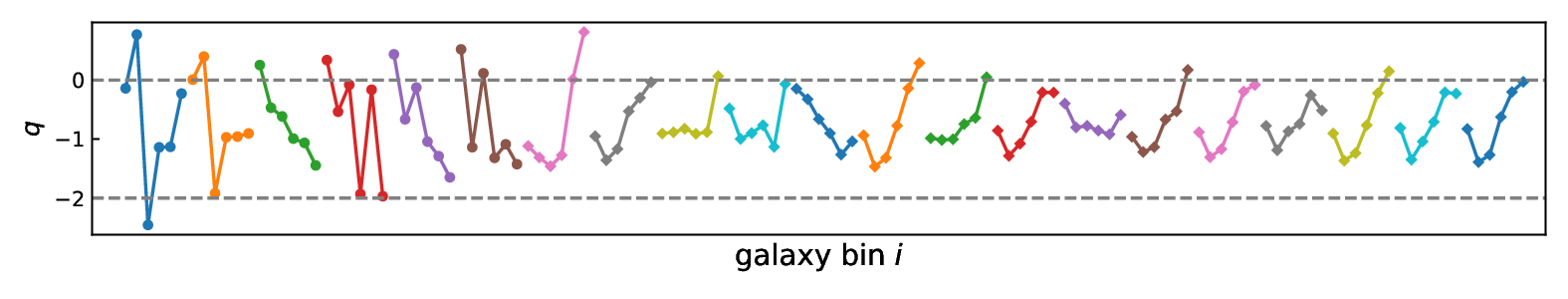}
\caption{ \label{fig:qval} 
The measured magnification response $q_i$ in cosmoDC2 galaxy sub-samples within photometric redshift $0.8<z_{\rm ph}<1.2$. The galaxy bins are ordered by flux bands $ugrizY$ (first 6 pieces with circle dots), and then the color bands (the latter 15 pieces with square dots).
}
\end{figure*}

\begin{figure*}
\includegraphics[width=0.9\textwidth]{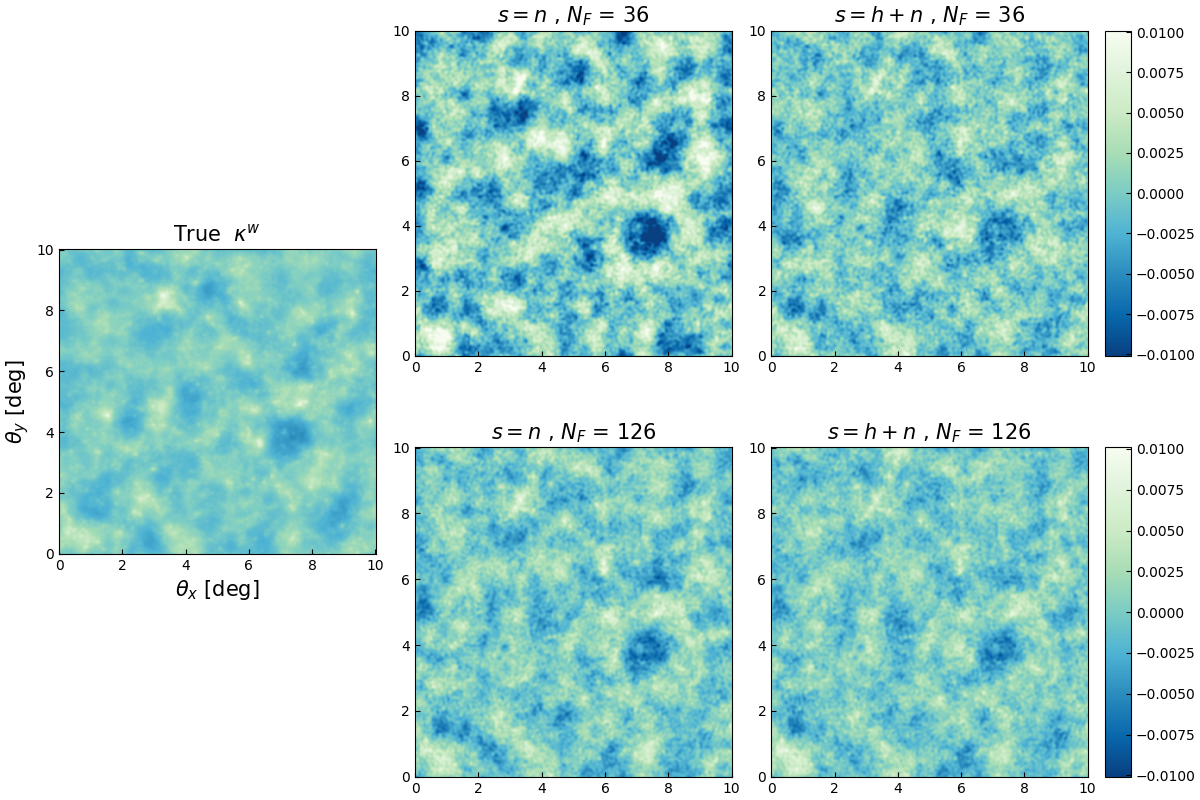}
\caption{ \label{fig:cILC_maps} 
The reconstructed $\kappa$ maps used cILC method compared with true $\kappa$ map. We adopt the same labels as in Fig.~\ref{fig:cILC_cross} for different schemes. We apply the wiener filter to suppress the shot noise after reconstruction for both the reconstructed maps and the true lensing map.
}
\end{figure*}
\begin{figure}
\includegraphics[width=1.0\columnwidth]{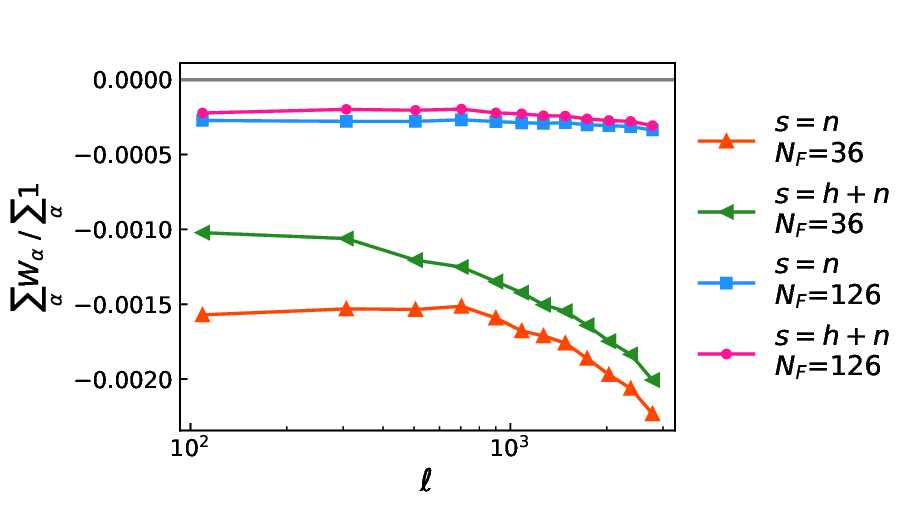}
\caption{ \label{fig:cILC_weisum}  
The mean value of the weights derived in cILC method. We adopt the same labels as in Fig.~\ref{fig:cILC_cross} for different schemes.
}
\end{figure}

\begin{figure*}
\includegraphics[width=0.95\textwidth]{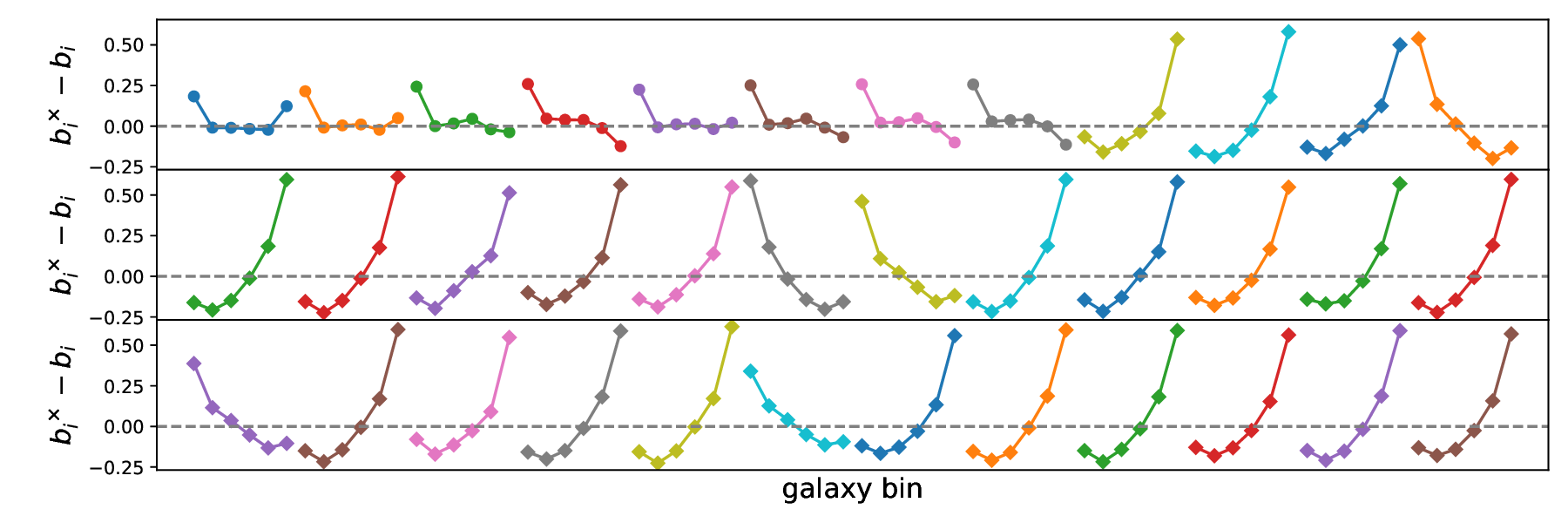}
\caption{ \label{fig:galaybias_diff}
The difference between galaxy bias $b$ and $b^\times$ measured in on the TNG300-1 galaxy samples at $\ell=700$,
The galaxy bins are ordered by flux bands $UBVKgriz$ (first 8 pieces with circle dots), and then the color bands (the latter 28 pieces with square dots).
}
\end{figure*}
\begin{figure}
\includegraphics[width=0.85\columnwidth]{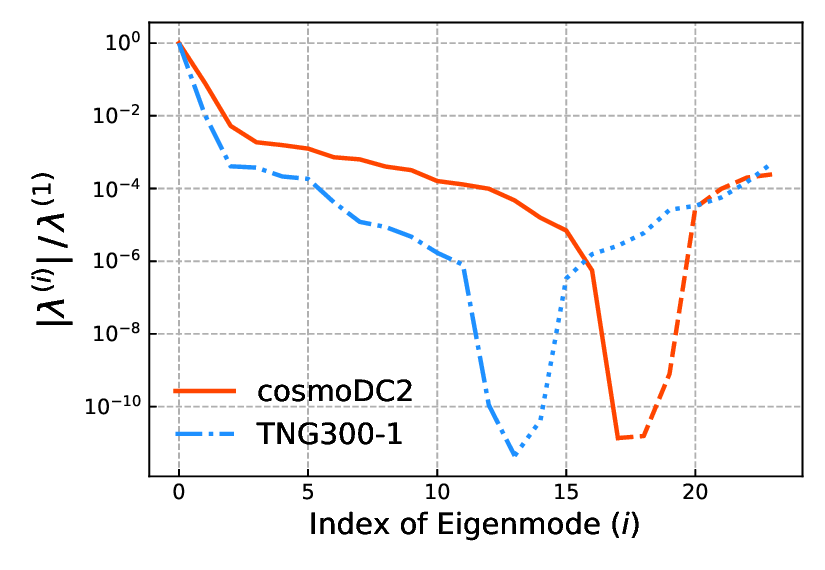}
\caption{ \label{fig:eigenval_griz} 
Eigenvalues distribution in the cosmoDC2 $\&$ TNG300-1 galaxy samples at $\ell=700$. Here we use the $griz$ bands' information for both two galaxy sets, therefore there are total $24$ bins. The eigenvalues are normalized by the first one, and we also take the absolute values of $\lambda^{(i)}$ due to the negative eigenmodes. The $\lambda^{(i)}$ values after the turning point are negative.
}
\end{figure}
\begin{figure*}
\includegraphics[width=0.85\textwidth]{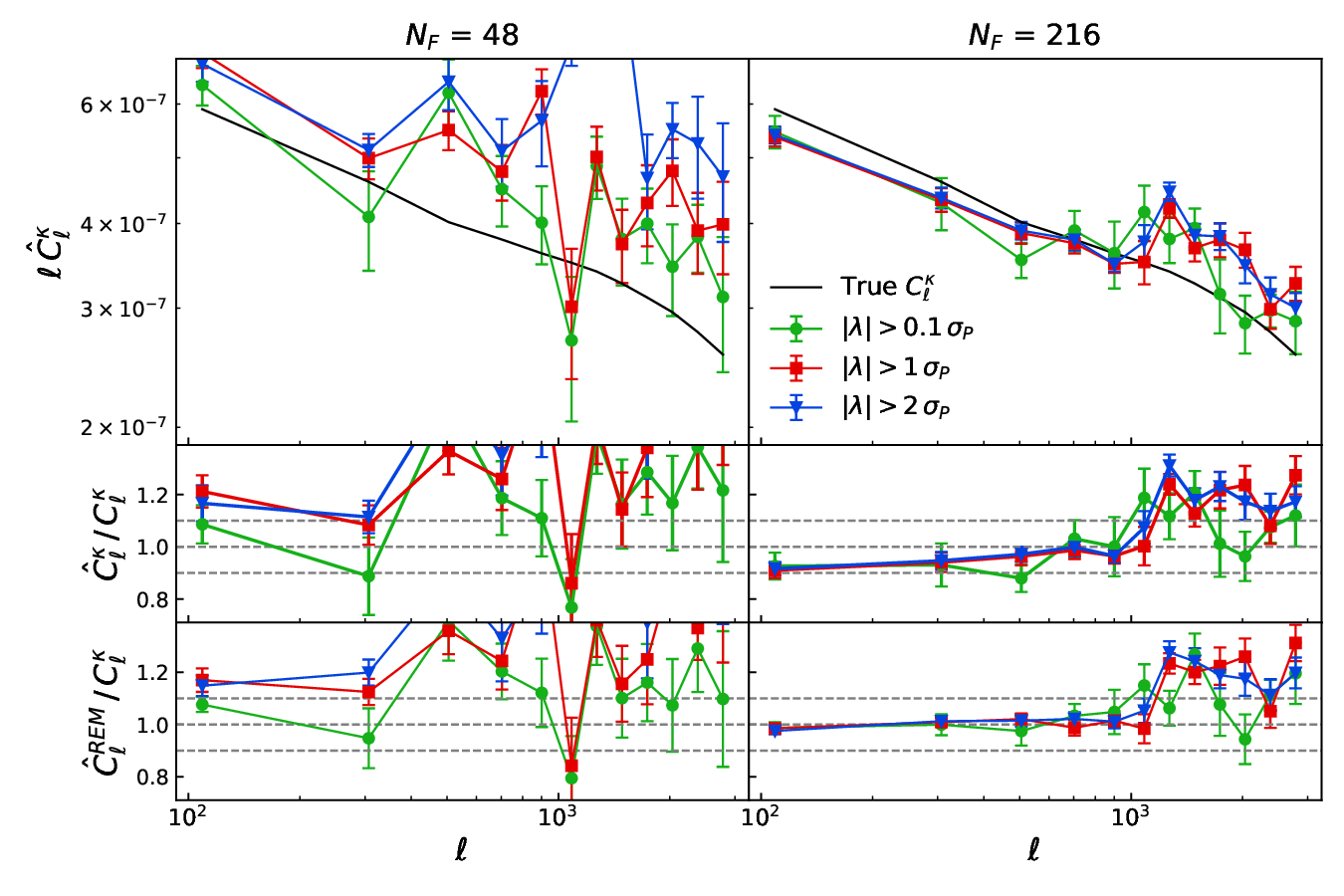}
\caption{ \label{fig:recont_ABS-tng300} 
ABS reconstruction performance on TNG300-1 galaxy samples. 
Same as Fig.~\ref{fig:recont_ABS}.
}
\end{figure*}
\begin{figure*}
\includegraphics[width=0.85\textwidth]{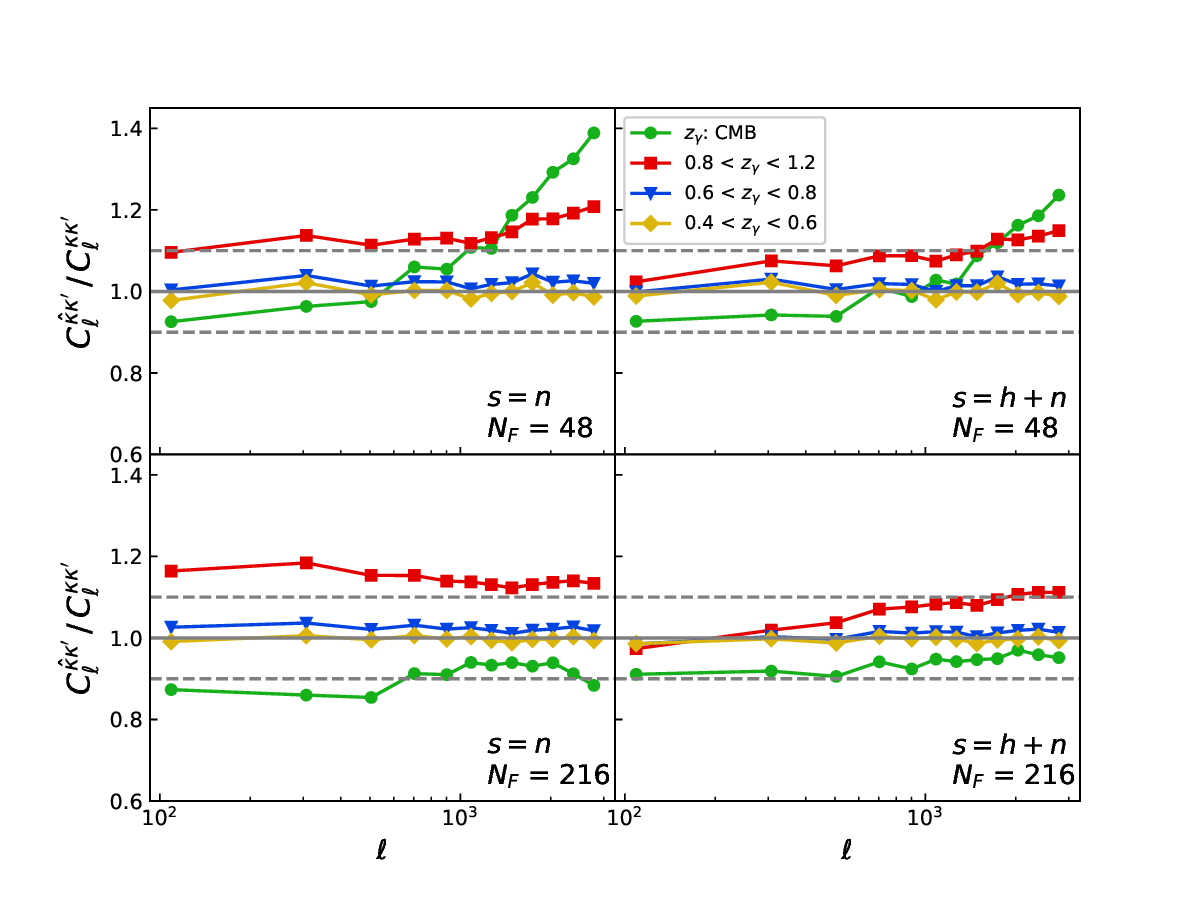}
\caption{ \label{fig:cILC_cross-tng300} 
cILC reconstruction performance on TNG300-1 galaxy samples. 
Same as Fig.~\ref{fig:cILC_cross}.
}
\end{figure*}
\begin{figure}
\includegraphics[width=1.0\columnwidth]{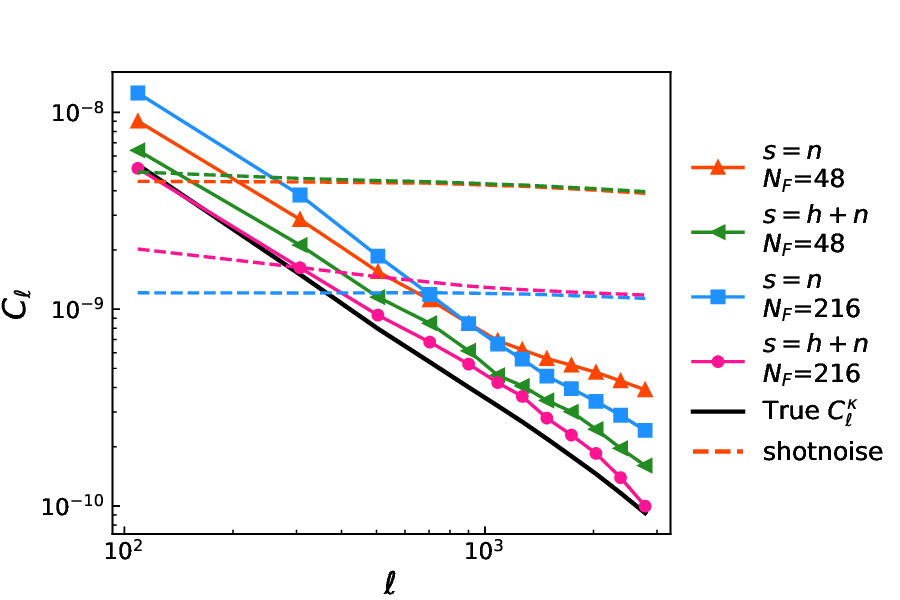}
\caption{ \label{fig:cILC_auto-tng300} 
cILC reconstruction performance on TNG300-1 galaxy samples. 
Same as Fig.~\ref{fig:cILC_auto}.
}
\end{figure}


\section{Data Construction on Simulation}
\label{sec:appendix_samples}

\subsection{Galaxy Samples}
\label{sec:appendix_samples_1}

We utilize two sets of galaxy samples, cosmoDC2 and TNG300-1, with corresponding dark matter distribution for performance test. 
We perform the fiducial analysis based cosmoDC2 samples, and additionally, we present the results based on TNG300-1 samples in Appendix~\ref{sec:tng300-results} to make the conclusions robust.

The cosmoDC2 is a synthetic galaxy catalog generated for LSST survey \cite{korytov2019cosmodc2}, and it provides a lightcone covering $440\,{\rm deg}^2$ of sky area, with $\sim2.6$ billions of galaxies distributed to redshift $z=3$. The magnitude $ugrizY$ photometric bands and the host halo are provided for the galaxies. 
The hydrodynamic simulation TNG300-1 is a publicly available simulation from the IllustrisTNG project \cite{Springel_2017, Nelson_2017, Pillepich_2017, Naiman_2018, Marinacci_2018}. IllustrisTNG is a suite of large volume, cosmological, gravo-magnetohydrodynamical simulations run with the moving-mesh code ${\tt APEPO}$ \cite{springel2010pur}. IllustrisTNG implements a set of physical processes to model the galaxy formation process including gas cooling, star formation and evolution, supernova feedback \cite{pillepich2018simulating}, and AGN feedback \cite{weinberger2016simulating}. This project includes TNG50, TNG100 and TNG300, with boxsize of $50$ Mpc, $100$ Mpc and $300$ Mpc respectively. We choose TNG300-1 for our analysis for its largest amount of galaxy samples. We identify all luminous subhalos in the TNG300-1 as galaxies.

However, the accessed cosmoDC2 products do not included the associated 3D matter field, which is based on the Outer Rim simulation \cite{Heitmann_2019}. Regarding the TNG300-1 simulation, the products we assess are complete but the boxsize $L\,=\,205\,{\rm Mpc/h}$ is limited for our analysis, which lacks the statistic for linear scale $\ell\sim 100$. To solve these problems, we combine the galaxies from cosmoDC2$\,\&\,$TNG300-1 with dark matter simulation CosmicGrowth \cite{jing2019cosmicgrowth}. The CosmicGrowth simulation adopts a flat WMAP cosmology \cite{hinshaw2013nine}, of which the boxsize is $L\,=\,1200\,h^{-1}{\rm Mpc}$ and particle number is $3072^3$. The cosmological parameters adopt $\Omega_b=0.0445\,,\;\Omega_c=0.2235\,,\;h=0.71$ and $\sigma_8=0.83$. We use the galaxy-host halo relation of the galaxy in cosmoDC2$\;\&\;$TNG300-1 to assign galaxies to the CosmicGrowth simulation halos, preserving the conditional probability of galaxy contents given hosthalo mass \cite{korytov2019cosmodc2}. Because of the distinction of these simulations resulted from different cosmological parameters and the simulation details, we do an abundance matching between cosmoDC2$\;\&\;$TNG300-1 and CosmicGrowth halos, and then assign galaxies in cosmoDC2$\;\&\;$TNG300-1 to CosmicGrowth halos.


\subsection{Sub-Samples Definition}
\label{sec:appendix_samples_2}

We generate 144 light-cones of galaxy and corresponding dark matter distribution for cosmoDC2$\;\&\;$TNG300-1, with angular size $10^\circ\times10^\circ$. Since we have only one realization, we divide the snapshot box into slices with $\Delta z\simeq0.04$, and then rotate, flip and shuffle these slices before stacking them into 3D light-cones. We apply the magnitude cut $\{\,26.1,\,27.4,\,27.5,\,26.8,\,26.1,\,24.9\,\}$ on the $ugrizY$ bands respectively for the cosmoDC2 galaxies, corresponding to the expected depth of coadded images from LSST \cite{ivezic2019lsst}. For the TNG300-1 galaxies, we only apply the same magnitude cut on $griz$ bands.
We choose the photometric scatter $\sigma_z=0.05(1+z)$ to define the observed photometric redshift, and then apply the photometric redshift cut $0.8<z_{\rm ph}<1.2$ to obtain the fiducial galaxy sample for performance test, where the surface density is $\bar{n}_g= 11.8\,\&\,15.7\,{\rm arcmin}^{-2}$ for cosmoDC2$\;\&\;$TNG300-1 respectively. We also apply other redshift cuts to get the redshift distribution as shown in Fig.~\ref{fig:cILC_cross}. We then project these galaxy samples into flat plane, and then assign the samples into 2D mesh using CIC method with grid number $1000\times1000$. The corresponding galaxy lensing map and the CMB lensing map (source $z_s=1100$) are generated using Born approximation in the procedure. 

We generate the sub-samples with equal galaxy number in each galaxy bin. For the cosmoDC2 samples, we divide the galaxy into sub-samples according to 6 flux bands $ugrizY$ and the 15 galaxy color band $g-r,\,r-i,\cdots$, all the combination of flux bands. There are 6 sub-samples for each galaxy property, therefore total 126 sub-samples for cosmoDC2 galaxy. For the TNG300-1 samples, the available flux bands are $UBVKgriz$, therefore total 216 sub-samples for TNG300-1 galaxy. We present the galaxy redshift distribution for all the sub-samples in Fig.~\ref{fig:nz_cosmoDC2} $\&$ \ref{fig:nz_tng300}. It shows that the redshift evolution is significant for cosmoDC2 galaxies, leading to the dispersion of the sub-samples distribution, especially for the galaxy color evolutions. But for the TNG300-1 samples, there are little impacts on the sub-samples distribution, and particularly the dispersion is minor for the flux divided samples. 

On the one hand, the variation among the sub-samples galaxy distribution profiles intensifies the difference between the galaxy bias $b_i$ and $b_i^\times$ apart from the intrinsic redshift evolution, as mention in the main text. On the other hand, it sources the difference of weak lensing effect for sub-samples. 
In principle, we should take account it for each sub-sample. However, we adopt the same $\kappa$ field times the constant magnification response $q_i$ for the sub-samples overdensity (using Eq.~(\ref{equ:deltagL})), for the reason: 
\begin{itemize}
\item[(a)] We can derive the same conclusions from the results based on the TNG300-1 galaxy samples, which is impacted little by the dispersion in the sub-samples distribution. It is a validation that our simplifying is a reasonable approximation. 
\item[(b)] In the realistic galaxy data, with prior knowledge of galaxy evolution and their redshift distribution, we can apply appropriate redshift cut to match the redshift distribution of each sub-sample. Therefore it is not an essential problem and is able to solve with the improvement of measuring redshift. 
\end{itemize}
Theoretically speaking, a galaxy is lensed only by the matter distribution in the front of it, and the response $q\equiv q(z)$ is the function of redshift \cite{wenzl2023magnification, Challinor_2011}. In linear order, the magnification contribution is
\begin{eqnarray}
\Delta^{\rm mag} =  \int{ q(z_s)n_g(z_s)W_\kappa(\chi;z_s)\delta_m(z) \, d\chi dz_s }   \;\;. 
\end{eqnarray}
where $W_\kappa(\chi;z_s)$ is the thin lensing kernel with lensing source at $z_s$. The intrinsic variation among the sub-samples magnification is accompanied by the $q(z_s)$ redshift dependence, determined by this equation. The complete and self-consistent treatment of the evolution effect exceeds the scope of this article. Meanwhile, the redshift resolution of our galaxy light-cones does not allow us to implement such sophisticated manipulation. Therefore we adopt the simplification of Eq.~(\ref{equ:deltagL}).


\subsection{Magnification Response}
\label{sec:appendix_samples_3}

We measure the galaxy magnification response in the simulated galaxy samples directly. Due to the multi-cuts on the galaxy flux bands and the additional color division we adopt, the $\alpha$ in the magnification response expression $q\equiv2(\alpha-1)$ is not simply the logarithmic slope of galaxy luminosity function. Before we make the magnitude cut on the selected galaxy samples, we alter the intrinsic galaxy magnitude by 
\begin{eqnarray}
m^L = m^{\rm intri} - 2.5\log_{10}(1+2\kappa)
\end{eqnarray}
where $\kappa$ is the lensing convergence value interpolated from the weak lensing map constructed in simulation. Then we apply the magnitude cut and obtain the lensed galaxy, and calculate its overdensity map $\hat{\Delta}_g^{L\prime}$. We also calculate the unlensed galaxy overdensity map $\hat{\Delta}_g$, and the difference of two is expected to be the lensing convergence multiplied by $2\alpha$ value. We calculate the cross correlation between their difference and true weak lensing map
\begin{eqnarray}
\hat{C}^\times \equiv \langle\, (\hat{\Delta}_{g,\ell}^{L\prime} -\hat{\Delta}_{g,\ell})\,\kappa_\ell^* \,\rangle   \quad,
\end{eqnarray}
and estimate the $2\alpha$ factor by minimizing 
\begin{eqnarray}
\chi^2 = \sum_{88<\ell<1270} {  \left( \hat{C}^\times - 2\alpha C^\kappa\right)^2  \over  \sigma_\ell^2 }  \quad,
\end{eqnarray}
here $\sigma_\ell^2$ is the variance of $\hat{C}^\times$. The observed area element magnification effect is imported by simply subtracting $2$, and finally we obtain the magnification response values $q=2(\alpha-1)$ for the performance test. The results of estimated $q$ are shown in Fig.~\ref{fig:qval}.


\section{Ancillary Results}

\subsection{cILC Reconstruction}

We present the reconstructed $\kappa$ using cILC method in Fig.~\ref{fig:cILC_maps}. Since the reconstruction is dominated by the shot noise, we apply the wiener filter $w^{\rm WF}_\ell = C^\kappa_\ell /\left(C^\kappa_\ell + N\right)$ to reveal the clustering feature above the noise dominating scale, where $C^\kappa_\ell$ is the true lensing power spectrum and $N= 4\pi f_{\rm sky} /\bar{n}_g$ is the constant shot noise of the galaxy samples.
All the reconstructed map is noisy and blurred, especially the scheme using Eq.~(\ref{equ:cILC_condi2}) with only flux information due to the residual matter clustering contamination. Both the inclusion of color information or higher order eigen-components increase the similarity between the reconstruction and true one, and significantly suppress the residual matter clustering. The scheme using Eq.~(\ref{equ:cILC_condi1}) with flux $\&$ color information achieve remarkable recovery of the most features after wiener filtering, but the large scale feature is weakened due to lensing power leaked. The conclusions are same as in the main text, since Fig.~\ref{fig:cILC_maps} is an equivalent description with Fig.~\ref{fig:cILC_auto}.

Previous works have shown that the constraint condition
\begin{eqnarray}
\label{equ:cILC_oldconst}
\sum_i w_i = 0
\end{eqnarray}
to require the mean galaxy bias values vanishing, rather than Eq.~(\ref{equ:constrains2}), is a sub-optimal but practical approach in realistic survey data \cite{hou2021weak,qin2023weak}. We present the averaging cILC weight in Fig.~\ref{fig:cILC_weisum}, recovering the finding in Ref.~\cite{ma2024method}. More interestingly, among these four schemes, the scheme with better suppression of the additive bias presents more consistent with the condition $\langle W\rangle\equiv\sum_iW_i/\sum_i 1\simeq0$. It further confirms that the constraint Eq.~(\ref{equ:cILC_oldconst}) is reasonable to analyze the real galaxy catalog.


\subsection{TNG300-1 results}
\label{sec:tng300-results}

To make the conclusion robust, we repeat the analysis based on the TNG300-1 galaxy samples, and present the results in Fig.~\ref{fig:recont_ABS-tng300}, Fig.~\ref{fig:cILC_cross-tng300} $\&$ Fig.~\ref{fig:cILC_auto-tng300}. For simplicity, we do not add any observation correction such as dust extinction \cite{Nelson_2017}, but nevertheless it is a useful cross-validation. We present the intrinsic difference between galaxy bias and galaxy-lensing cross-correlation bias in Fig.~\ref{fig:galaybias_diff}.

Due to the intrinsic properties of TNG300-1 galaxy and the lack of survey effects which differ from band to band, the galaxy clustering signals separated by different bands share high similarity and correlation. It means that there are fewer effective degrees of freedom in the cross band power matrix of TNG300-1 samples compared to cosmoDC2. We reveal it using the eigenvalue distribution, shown in Fig.~\ref{fig:qval}. The eigenvalues of TNG300-1 drop faster than those of cosmoDC2, though both tracing the identical matter distribution, which means that there are fewer independent components in TNG300-1 galaxy samples compared to cosmoDC2 samples. 
\textcolor{black}{
Two direct consequences of the fewer independent components in ABS reconstruction are shown in Fig.~\ref{fig:recont_ABS-tng300}. 
One is the worse performance of the reconstruction using only flux information, which is unstable even in the linear region. Although the reconstructed results capture the cosmic magnification with the same order of magnitude, they are dominated by the galaxy clustering contamination and the noise fluctuation. 
Another consequence is the shot noise domination of the small scale for the reconstruction with flux$\,\&\,$color information. There are much more tiny eigenmodes in the steep eigenvalue distribution of TNG300-1 samples, and they are overwhelmed by the shot noise. We have to exclude these unstable eigenmodes, while partial lensing signal is excluded simultaneously. The loss of lensing signal seriously impacts on the noise-dominatng small scale, and results in the systematic overestimation at scale $\ell\gtrsim 1000$ for threshold $\lambda_{\rm cut}\gtrsim 1\sigma_P$.
Nevertheless, the ABS results based on TNG300-1 confirms the conclusions in the main text. 
}

The cILC reconstruction based on the TNG300-1 samples also derives the similar results as in the main text, shown in Fig.~\ref{fig:cILC_cross-tng300} $\&$ \ref{fig:cILC_auto-tng300}. A slight difference appears in the scheme including color information, which show minor improvement compared to those without color information. We suppose the domination in the reconstruction noise is the galaxy stochasticity rather than the residual matter clustering, therefore it is ineffective to include more galaxy bins for suppressing the shot noise or canceling the matter clustering, but it is efficient to suppress high order eigen-components to clear the stochasticity noise. Apart from this, we recover the conclusions shown in the main text.


\bibliographystyle{apsrev4-2}
\bibliography{citations}

\end{document}